\documentclass[12pt]{article}
\usepackage{scicite}
\usepackage{times}

\usepackage{graphicx}
\usepackage{epsfig}
\newenvironment{sciabstract}{%
\begin{quote} \bf}
{\end{quote}}
\newcounter{lastnote}
\newenvironment{scilastnote}{%
\setcounter{lastnote}{\value{enumiv}}%
\addtocounter{lastnote}{+1}%
\begin{list}%
{\arabic{lastnote}.}
{\setlength{\leftmargin}{.22in}}
{\setlength{\labelsep}{.5em}}}
{\end{list}}

\title{Design of crystal-like aperiodic solids with\\ selective disorder--phonon coupling}

\author
{Alistair R. Overy,$^{1,2}$ Andrew B. Cairns,$^{1}$ Matthew J. Cliffe,$^{1}$\\
Matthew G. Tucker,$^{2,3}$ and Andrew L. Goodwin$^{1\ast}$ \\
\\
\normalsize{$^{1}$Department of Chemistry, University of Oxford, Inorganic Chemistry Laboratory,}\\
\normalsize{South Parks Road, Oxford OX1 3QR, U.K.}\\
\normalsize{$^{2}$Diamond Light Source, Chilton, Oxfordshire, OX11 0DE, U.K.}\\
\normalsize{$^{3}$ISIS Facility, Rutherford Appleton Laboratory, Harwell Oxford, Didcot,}\\
\normalsize{Oxfordshire OX11 0QX, U.K.}\\
\\
\normalsize{E-mail: andrew.goodwin@chem.ox.ac.uk.}
}
\date{}

\begin{document} 

\baselineskip24pt

\maketitle 

\begin{sciabstract}
Functional materials design normally focuses on structurally-ordered systems because disorder is considered detrimental to many important physical properties. Here we challenge this paradigm by showing that particular types of strongly-correlated disorder can give rise to useful characteristics that are inaccessible to ordered states. A judicious combination of low-symmetry building unit and high-symmetry topological template leads to aperiodic ``\emph{procrys\-talline}'' solids that harbour this type of topological disorder. We identify key classes of procrystalline states together with their characteristic diffraction behaviour, and establish a variety of mappings onto known and target materials. Crucially, the strongly-correlated disorder we consider is associated with specific sets of modulation periodicities distributed throughout the Brillouin zone. Lattice dynamical calculations reveal selective disorder--phonon coupling to lattice vibrations characterised by these same periodicities. The principal effect on the phonon spectrum is to bring about dispersion in energy rather than wave-vector, as in the poorly-understood ``waterfall'' effect observed in relaxor ferroelectrics. This property of procrystalline solids suggests a mechanism by which strongly-correlated topological disorder might allow new and useful functionalities, including independently-optimised thermal and electronic transport behaviour as required for high-performance thermoelectrics.
\end{sciabstract}

The relationship between building-block geometry and bulk material structure is one of the cornerstones of structural science. By way of example, the solid phases of elemental Xe \cite{Natta_1930}, C$_{60}$ \cite{David_1991} and human adenovirus \cite{Reddy_2010} are all structurally related, not by virtue of any particular chemical similarity but because each of these phases reflects the same solution to the problem of packing weakly-interacting spherical objects in three-dimensional (3D) space. The reticular approach to understanding zeolite and metal--organic framework (MOF) topologies is related, because it links the geometries of molecule-like components (coordination polyhedra, molecular linkers) to the 3D architectures formed by their assemblies \cite{Meier_1968,Eddaoudi_2001}. The importance of structure in determining the physical properties of solids is what then gives sense to the approach of developing new types of functional materials through informed design of their constituent building blocks. It is precisely this type of ``ground-up'' approach that has recently been exploited in the rational design of \emph{e.g.}\ solid oxide fuel-cell cathodes and room-temperature multiferroic candidates \cite{Dyer_2013,Pitcher_2015}.

To a large extent, the rich structural information accessible using crystallographic techniques has focussed effort on the design of crystalline materials. There are obvious functional advantages to the long-range periodicity characteristic of crystals, because it governs useful \emph{correlated} properties---including the lattice dynamics, electronic states, and charge, orbital and magnetic order. Moreover crystal symmetry is central to mechanical properties such as piezoelectricity and ferroelasticity, and is clearly pivotal in determining phase transition behaviour. While certain building block geometries allow---or can even force \cite{Steinhardt_1996,Damasceno_2012}---non-crystalline assemblies, the link to function is usually much less clear in these cases. Indeed the received wisdom is that disorder is something to be avoided, despite increasingly strong empirical evidence that links disordered states to advanced functionalities \cite{Keen_2015}. So the development of disorder--property relationships, and the eventual control over these properties through suitable building-block design have emerged as key challenges in the field.

Here we develop an approach of intentionally designing functional disordered materials by focussing on systems in which structural disorder is extremely strongly correlated. The type of disorder we consider is similar to that found in ice, and our paper begins by developing a generalisation of ice-like states to arbitrary materials geometries. We proceed to establish a link between the geometry of structural building blocks and the propensity for specific types of strongly-correlated disorder in the resulting material assembly. This is the design element of our approach. Having made this connection, we develop a geometric basis for describing correlated disordered states. Our paper concludes by demonstrating how the correlated disorder deliberately engineered within one such material affects its lattice dynamics in a highly specific manner. This is the functional element of our approach because the effect we observe suggests, for example, a fundamentally new way of optimising thermoelectric response.

Our starting point is the simple toy model of square ice, in which water molecules are arranged on a square lattice and oriented so as to satisfy sensible hydrogen-bonding rules [Fig.~1(a)]. There being no unique way of satisfying these rules the system is disordered, and---like the real-world examples of cubic ice (I$_{\rm c}$) and its nano-confined variants \cite{AlgaraSiller_2015}---is characterised by a degenerate manifold of structural ground-states \cite{Pauling_1935}. An idea we will come to develop is that this propensity for disorder is encoded in the combination of the symmetry of the water molecule (\emph{i.e.}\ the structural building block) and the lattice on which the water molecules are arranged (here enforced by the directionality of the chemical interactions between building blocks). Any system that shares these geometric features will be characterised by the same configurational degeneracy. So, for example, replacing O--H$\ldots$O linkages by the M--C--N--M motif found in transition-metal cyanides gives a mapping that---in 3D---relates head-to-tail cyanide disorder in Cd(CN)$_2$ to water molecule orientations in cubic ice [Fig.~1(b)] \cite{Fairbank_2012}. The question of O/N ordering within square-grid-layers of transition-metal oxynitrides presents a related problem, which maps onto the square ice model following geometry inversion from one site to the next [Fig.~1(c)] \cite{Camp_2012}. These examples involve compositional or orientational modulations of the square lattice, but the same ideas are well known to translate to a variety of modulation types, many of which are key to material function: \emph{e.g.}, displacive, electronic, charge density, spin density, orbital and spin orientation [Fig.~1(d)]. For ice-like disorder on the diamond lattice, these mappings are well established in the literature: hence the ``Coulomb phases'' \cite{Henley_2010} of charge \cite{Shoemaker_2010}, orbital \cite{Chern_2011}, and spin \cite{Bramwell_2001} ices.

In seeking to generalise ice-like states, we take our lead from the reticular chemistry approach for generating network structures \cite{Wells_1984,Yaghi_2003}. The key idea here is that lattice topologies can be considered in terms of the assembly of nodes and linkers [Fig.~2(a)]: the geometry of the node determines the possible topologies of the corresponding lattice. In this context, the various square-ice systems of Fig.~1 can be considered perturbations of the square lattice in which two adjacent linkers are distinguished amongst the four that meet at each square node. If we identify three linkers rather than two, then we generate a distinct family of disordered configurations that---from a reticular chemistry perspective---might be considered to arise from the linking of T-shaped building blocks [Fig.~2(b)]. As for the square ices, there are many possible realisations of this same state: one mapping is to Anderson's resonance valence bond (RVB) description of singlet pair formation in cuprate superconductors \cite{Anderson_1987}; another is to so-called ``domino'' tilings of the plane \cite{Kasteleyn_1961}. There are in total just six cases to be considered for perturbations of the square lattice. Two of these are trivial (distinguishing either four or zero linkers); two are related to one another (distinguishing one linker being the same as distinguishing three); and the two cases that remain distinguish different pairings of the four linkers, as shown in Fig.~2(c,d). We have already met the first of these cases in the guise of square ice [Fig.~1(c)]; the second case---in which the linkers distinguished are opposite one another---is ordered and results in symmetry breaking of the underlying square lattice. So there is a nontrivial relationship between perturbations of the node symmetry and the resulting configurational degeneracy.

Ice-like configurations are not confined to perturbations of the square lattice. Equivalent states for the hexagonal, triangular, diamond, and cubic nets are enumerated in Fig.~2(e)--(s); those for the pyrochlore lattice are included as SI. The extent of disorder can be deduced from the corresponding diffraction patterns, which contain structured diffuse scattering in cases where there is strong correlated disorder [Fig.~2] \cite{Keen_2015}. What emerges is that a substantive fraction of these systems admit large configurational entropies, with a complex relationship between node geometry and extent of correlated disorder. So as to provide insight into this relationship, we generalise Pauling's approximation for the configurational entropy of ice \cite{Pauling_1935,Camp_2012}:
\begin{equation}
S_{\rm{config}}\simeq R\ln\left(\frac{n}{2^{d/2}}\right).
\end{equation}
Here $d$ represents the underlying node connectivity and $n$ corresponds to the number of symmetry-equivalent node perturbations of a given type. The value of $n$ can often be determined by inspection, but it is given more rigorously by the ratio of the orders of the point groups of the parent and perturbed node geometries; for example, $n=|D_{4h}|/|C_{2v}|=4$ for square ice. We call $p=n/2^{d/2}$ the Pauling number, with the significance that maximising $p$ maximises the propensity for disorder. In this way one expects low-symmetry perturbations of high-symmetry lattices to lead to states of the greatest configurational entropy, a qualitative relationship that is borne out in practice [Fig.~2].From a materials design perspective, what we are saying is that building block geometry and the arrangement of the interactions between building blocks can together encode for specific types of correlated structural disorder.

Common to many of the various configurations of Fig.~2 is the absence of translational periodicity characteristic of the crystalline state. For a given configuration, every node experiences the same local environment---and hence it is not meaningful to think of these structures as defective in the vernacular sense. Yet there is no unit cell and space group symmetry that properly describes the topological connectivity. We proceed to argue that these systems should not be considered as crystals, but form a separate class of aperiodic solid with its own characteristics. We will use the term \emph{pro}crystalline to describe this state and to emphasise that conventional crystals might be seen as a special case of the definitions that follow. The procrystalline state is a dense, overlapping packing of identical fundamental structural units (we use the term ``\emph{neighbourhoods}''), which are positioned periodically but orientationally permuted as permitted by the point symmetry of the neighbourhood geometry. For magnetic systems, these permutations may involve time reversal operations as realised in \emph{e.g.}\ the Ising spin ices \cite{Bramwell_1998}. In simple cases the neighbourhood corresponds to the Dirichlet-Voronoi cell of the underlying lattice augmented to include neighbouring, correlated lattice points [Fig.~2(b)] (see SI for further discussion). Whereas crystals correspond to the special case in which the neighbourhood orientations are themselves periodic, the more general procrystalline state allows for discrete orientational disorder (\emph{cf}.\ the continuous orientational degrees of freedom in \emph{e.g.}\ plastic crystals and superionics). Any such disorder will always be correlated since neighbourhoods overlap. Because the underlying neighbourhood lattice is periodic, all procrystals admit a Bragg diffraction pattern and have a well-defined reciprocal lattice. This diffraction pattern can be analysed using conventional crystallographic approaches but doing so yields a structural model in which neighbourhoods are averaged over their different possible orientations and all information regarding orientational correlation is lost; for example, the states represented by panels (r) and (s) in Fig.~2 share \emph{identical} Bragg diffraction patterns in spite of their distinct local symmetries. Like crystals, procrystalline phases are characterised by macroscopic point symmetry that can be as high as that of the neighbourhood lattice. Yet, unlike crystals, they can support a complete absence of any point or translational symmetry at the microscopic level. It is the existence of a periodic 3D reciprocal lattice that distinguishes procrystals from incommensurate and quasicrystalline phases, and which also guarantees a well-defined Brillouin zone (BZ) and Bloch-wave description of phonon and electronic states.

The structures of a number of well- and lesser-known materials can be thought of in precisely these terms. This is true by construction for any phase with ice-like disorder; in addition to the various systems described above, the family of ferroelectric phases related to KH$_2$PO$_4$ is an obvious additional example \cite{Slater_1941}. Similarly well established are the statistical mechanical models of RVB \cite{Anderson_1973} and loop \cite{Kondev_1997} states, which correspond to procrystalline lattices with, respectively, one and two linkers distinguished for each node. Hence physical realisations of either class also fall under our definition (\emph{e.g.} TaS$_2$ \cite{Tosatti_1976} and SrTaO$_2$N \cite{Gunther_2000}). A less obvious example is the assembly of \emph{p}-terphenyl-3,5,3$^\prime$,5$^\prime$-tetracarboxylic acid (TPTC) molecules on pyrolytic graphite to form a hydrogen-bonded network related to the procrystalline lattice illustrated in Fig.~2(e) [Fig.~3(a,b)] \cite{Blunt_2008}. This arrangement maps onto the so-called ``rhombus'' or ``lozenge'' tiling, which in turn corresponds at once to both the Ising triangular antiferromagnet and the RVB description of $\pi$-bonding in graphene \cite{Wannier_1950,Fisher_1961,Kasteleyn_1963}. These equivalences are straightforwardly seen in reciprocal space: Fourier transform of the scanning tunnelling microscopy image of Fig.~3(b) reveals the same distribution of diffuse scattering and ``pinch point'' features expected from our simple geometric model [Fig.~3(c,d)]. Indeed we expect the link between characteristic diffuse scattering patterns and particular procrystalline states to aid in diagnosing and understanding a range of problems of correlated disorder \cite{Keen_2015,Welberry_2015}. For powder diffraction measurements, often the only signature of this diffuse scattering is the presence of $hkl$-dependent anisotropic peak shape broadening. This is the case, for example, in the scattering patterns of Pd(CN)$_2$ and Pt(CN)$_2$; a procrystalline structural model based on connected square-planar [M(C/N)$_4$] units provides the first convincing description of their diffraction behaviour [Fig.~3(e,f)] \cite{Hibble_2011}. In other cases, procrystalline states (if not necessarily recognised as such) have been inferred from the combination of (i) a disordered average structure, with (ii) clear signatures of local distortions that can persist only if suitably correlated. Examples include Jahn Teller distortions in the high-temperature orbital-disordered phase of LaMnO$_3$ \cite{Bozin_2006,Ahmed_2009} and the high-pressure amorphous phase of ZrW$_2$O$_8$ \cite{Keen_2007}. So there is good evidence that a variety of procrystalline phases do exist, even if their structures are difficult to interpret using established crystallographic approaches.

But what of the link between correlated structural disorder and function? In principle, the existence of a well-defined BZ allows coupling between the structural modulations that characterise the procrystalline state and other physical properties that depend on periodicity---\emph{e.g.}\ the lattice dynamics and electronic band structure \cite{Valla_2009}. We tested for coupling of this type using as our example a two-dimensional oxynitride lattice [Fig.~1(c)]. The idea was to set up a simple harmonic lattice dynamical model in which we assigned different equilibrium values and stiffnesses to M--O and M--N bond lengths, and again to O--M--O, N--M--O, and N--M--N bond angles, and then to determine the extent to which correlated compositional order affected the phonon spectrum. Conventional lattice dynamical calculations are designed for periodic (crystalline) structures, and so we made use of a supercell Monte Carlo lattice dynamical (MCLD) approach \cite{Dove_1986,Goodwin_2004b} in order to treat compositional disorder explicitly (see SI for further discussion). We first benchmarked our calculations by determining the ``mean-field'' phonon dispersion expected for an average of the different force constant values; we found essentially perfect agreement between our MCLD calculations and those obtained using the GULP program \cite{Gale_1997} [Fig.~4(a)]. For random distributions of equal numbers of O and N atoms, the basic phonon structure was similar to the mean-field case, with a slight redistribution of energies among phonon branches [Fig.~4(b)]. This result is in agreement with \emph{ab initio} molecular dynamics studies of the related problem of the lattice dynamics of configurational glasses \cite{Fang_2013}. By contrast, for the procrystalline arrangement there was a dramatic dispersion in energy of the optic branches that we observed \emph{only} for the specific subset of wave-vectors of the form $\langle\frac{1}{2},\xi\rangle^\ast$---precisely the family of modulation periodicities characterising the procrystalline state in this case [Fig.~4(c,d)]. The qualitative similarity to the ``waterfall'' phonons observed in thermoelectrics and relaxor ferroelectrics is striking, and suggests a plausible origin for the phenomenon in those systems \cite{Gehring_2000,Delaire_2011}.

So this is our key result: strongly-correlated structural disorder allows selective control over physical properties that depend on periodicity. The implication for systems where thermal and electronic conductivities are mediated by, respectively, phonons and electronic states localised in different regions of the BZ is that disorder--phonon coupling offers a means of selectively reducing thermal conductivity (inversely proportional to phonon bandwidth) without affecting charge transport behaviour. This is an attractive design strategy for developing next-generation thermoelectrics, and one that contrasts with the use of ``rattlers'' which are indiscriminant in their $\mathbf k$-space coupling \cite{Christensen_2008}.

Our reticular chemistry methodology suggests a number of synthetic routes for realising new classes of functional procrystalline solids. Metal--organic frameworks (MOFs) are an obvious platform, given (i) they offer the requisite control over building unit geometry and (ii) their energetics tend to be dominated by local interactions \cite{Cairns_2013,Cliffe_2014}. While there is reduced scope for coupling between structural disorder and electronic behaviour in these systems, porosity percolation will certainly be affected by disorder and may in turn govern sorption, mechanical and ion storage properties \cite{Wessells_2011,Cliffe_2015}. In more conventional inorganics, local symmetry lowering can be achieved by covalency effects (as in mixed-anion perovskites) or by first- or second-order Jahn Teller distortion (as in the chalcogenide thermoelectrics). Moreover, because our analysis is essentially geometric in nature, there is clear scope to extend these concepts to magnetic or electronic states, or indeed to the macroscopic scale. The recent demonstration that disordered metamaterials can show strong structural coupling to light scattering processes is an example relevant to the generation of modern photonics \cite{Riboli_2014}. Thinking beyond the ground-state properties of procrystals, we anticipate the existence of novel collective and hidden degrees of freedom that promise a rich physics of their own; \emph{e.g.}, topological excitations\cite{Bak_1985} and/or ``hidden order'' transitions between distinct local symmetries \cite{Toudic_2008}.

% Stacking faults, FeOF

\baselineskip24pt

\bibliography{arxiv_2015_dbd}

\bibliographystyle{Science}

\begin{scilastnote}
\item A.R.O., A.B.C., M.J.C. and A.L.G. gratefully acknowledge financial support from the EPSRC (Grant EP/G004528/2), the ERC (Grant 279705), and from the Diamond Light Source to A.R.O. This project has received funding from the European Union's Horizon 2020 research and innovation programme under the Marie Sklodowska-Curie grant agreement No.~641887 (Project Acronym: DEFNET-ETN). The authors are grateful to U. Waghmare (Bangalore), D. A. Keen (ISIS), W. J. K. Fletcher (Oxford) and J. A. M. Paddison (Georgia Tech.) for relevant discussions.
\end{scilastnote}

\clearpage

\noindent {\bf Fig.\ 1.} \\
{\bf Correlated disorder in square ice analogues.} (a) A configurational fragment of square ice. The water molecules are arranged on a square lattice and are oriented so as to satisfy local hydrogen-bonding rules: each molecule accepts two hydrogen bonds and donates two hydrogen bonds. There is no unique solution to satisfying these local constraints, and so the square ice state is configurationally disordered. (b) A square-planar transition-metal cyanide configuration that maps onto the ice-like state in (a). Here each metal cation (coloured green) is coordinated by two nitrogen atoms (blue) and two carbon atoms (grey) such that each N atom is opposite to an C atom. (c) Transition metal oxynitrides adopt a related structure \cite{Camp_2012}, in which square-grid layers consist of metal cations coordinated by two nitrogen atoms (blue) and two oxygen atoms (red). The topological equivalence to the square ice configuration can be seen by alternately considering O--M--O and N--M--N orientations for neighbouring metal centres (shaded regions) \cite{Kondev_1997}. (d) Displacive modulation of a square MO$_2$ lattice by in-plane [MO$_4$] rotations gives configurations that again map onto the square ice state. The correspondence relates displacements above and below the plane with, respectively, O and N atoms of the oxynitride configuration shown in (c).
\clearpage

\noindent {\bf Fig.\ 2.} \\
{\bf Reticular design approach to generating procrystalline networks.} (a) High-symmetry building blocks connect to form familiar two- and three-dimensional networks: square (indigo), hexagonal (blue), triangular (green), diamondoid (orange), and cubic (red). (b)--(s) Distinguishing different possible subsets of linkers for these high-symmetry lattices gives a variety of ordered and disordered states, which are grouped here according to the parent lattice. For each panel, the perturbed node geometry is shown in the top-left corner, followed immediately below by a representation of the Pauling number $p$, a qualitative indicator of the propensity for disorder. A representative network configuration is shown as the main image, with nodes coloured according to their orientation. One suitable projection of the corresponding X-ray diffraction pattern is given in the bottom-left corner (see SI for further details). For the configuration shown in panel (b), two overlapping neighbourhoods are outlined in black. The Dirichlet-Voronoi cell of the neighbourhood lattice is shown in red; the neighbourhood itself is generated by augmenting this cell to include connected latticed points. The asterisk in panel (l) indicates that a single enantiomer of the node geometry is used (\emph{cf}.\ panel (k)); this node is chiral when constrained to lie in two dimensions. Further discussion, including extension to the pyrochlore lattice, is given as SI.

\clearpage

\noindent {\bf Fig.\ 3.} \\
{\bf Representative physical realisations of procrystalline states.} (a) Molecules of \emph{p}-terphenyl-3,5,3$^\prime$,5$^\prime$-tetracarboxylic acid (TPTC) self assemble on pyrolytic graphite to form a hexagonal procrystalline network related to that represented in Fig.~2(e) \cite{Blunt_2008}. One possible arrangement is shown here (left), together with the corresponding topological representation of the type used in Fig.~2 (centre); molecule orientations map onto the ``missing'' linkers of the topological representation. Shown in background is the corresponding rhombus tiling, which shares the same configurational degeneracy. There is a further one-to-one correspondence to ground states of the triangular Ising antiferromagnet (right): opposite vertices of each rhombus are decorated with pairs of ``spin-up'' (white circles) and ``spin-down'' (black circles) states. The constraint that each triplet of neighbouring spins (red triangle) contains at least one spin-up and one spin-down state perturbs the node symmetry of the underlying hexagonal lattice in the same way as does the pattern of hydrogen- and covalent-bonding in the molecular assembly. (b) A scanning tunnelling microscopy image of the corresponding experimental state [Adapted from Ref.~\cite{Blunt_2008}]. (c) The Fourier transform of the image in (b), shown regions of structured diffuse scattering. (d) The strikingly similar scattering pattern calculated for the hexagonal procrystalline network of Fig.~2(d). (d) A procrystalline model for the structure of Pd(CN)$_2$ and Pt(CN)$_2$ based on the state represented in Fig.~2(p): the two possible orientations of square-planar M(C/N)$_4$ nodes are shown in indigo and gold. (f) Comparison of the Rietveld fit for this structural model of Pd(CN)$_2$ (red lines) and the experimental X-ray powder diffraction data of Ref.~\cite{Hibble_2011} (black points, $\lambda=1.54$\,\AA); tick marks indicate the positions of parent Bragg reflections and the difference (data--fit) is shown in blue (see SI for further discussion).

\clearpage

\noindent {\bf Fig.\ 4.} \\
{\bf Selective disorder--phonon coupling in a procrystalline network.} (a) Mean-field phonon dispersion curves for a two-dimensional oxynitride lattice, determined using MCLD (coloured bands) and GULP (black lines) approaches. The width of the coloured bands reflects the standard uncertainty of the MCLD phonon frequencies, which is non-zero as a result of the finite configurational ensemble size used for phonon calculations (see SI for further discussion). (b) The phonon dispersion determined using MCLD for configurations in which equal numbers of O and N atoms are distributed randomly across the O/N sites of the same oxynitride lattice. There is a slight redistribution of phonon frequencies relative to (a) but the standard uncertainties remain essentially unchanged. (c) A third set of phonon dispersion curves, again determined using MCLD but for configurations in which O and N atoms have been distributed according to the correlated disorder (procrystalline) model of Fig.~2(c). There is now substantive ``waterfall''-like phonon broadening along the X--M direction which is an order of magnitude larger than the intrinsic finite-ensemble effects. (d) The calculated single crystal scattering pattern for the procrystalline model of Fig.~2(c), demonstrating that this particular state is characterised by modulations that---like the phonon broadening---are localised along the X--M branch of the BZ.

\clearpage

\begin{center}
\includegraphics{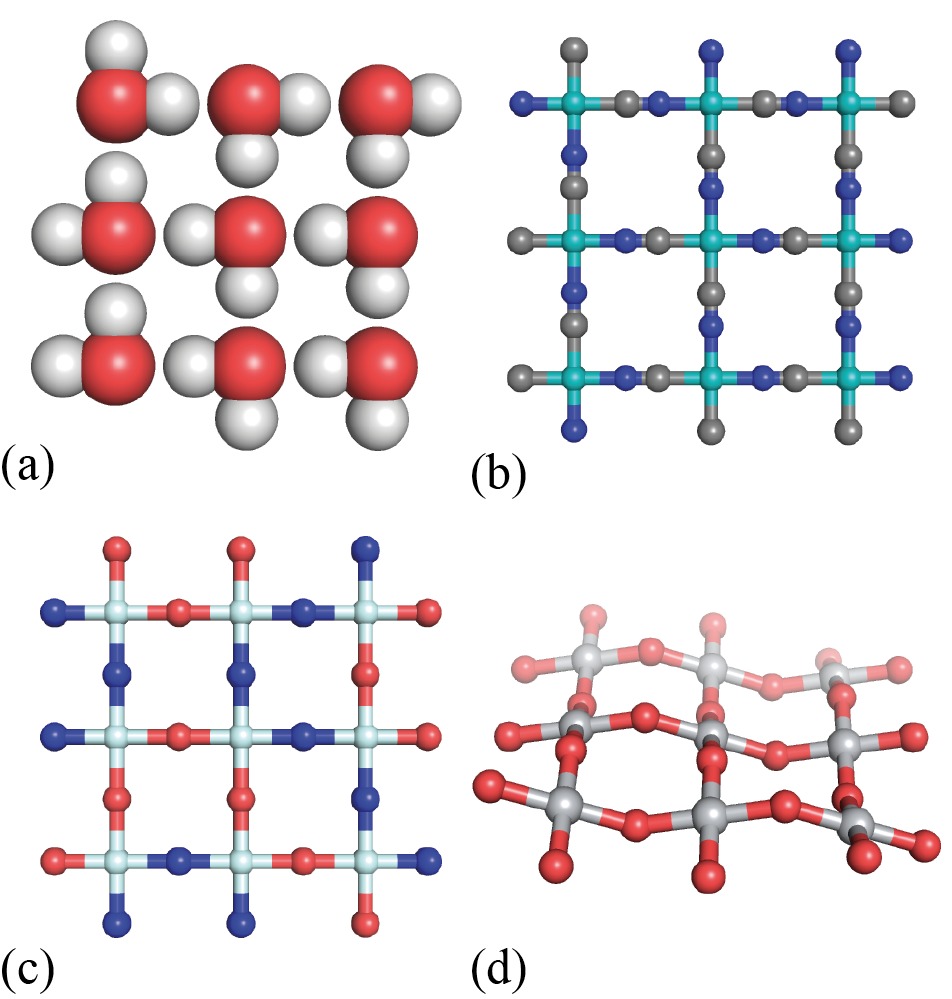}\\
FIGURE 1\\
\end{center}
\clearpage

\begin{center}
\includegraphics[width=\textwidth]{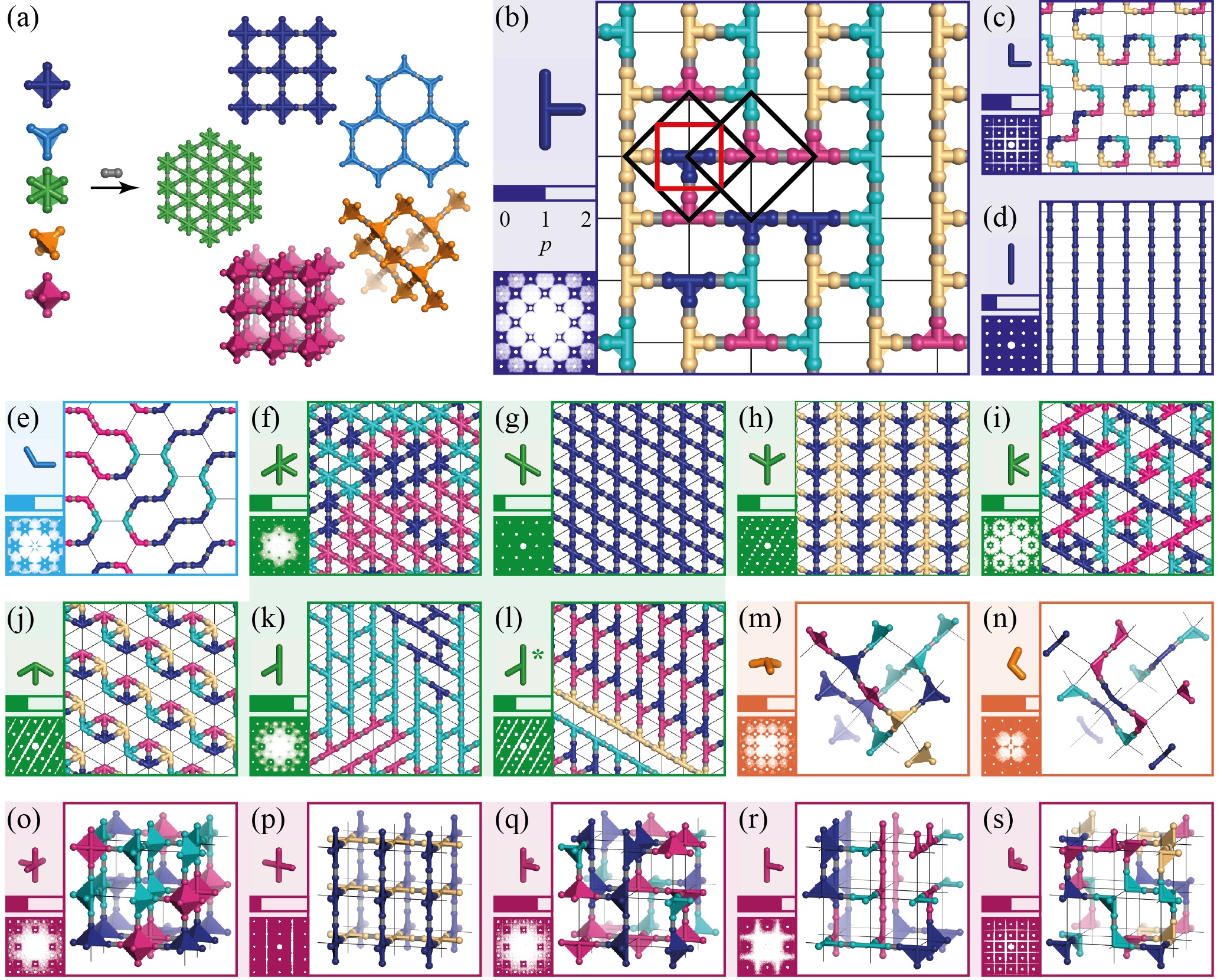}\\
FIGURE 2\\
\end{center}
\clearpage

\begin{center}
\includegraphics{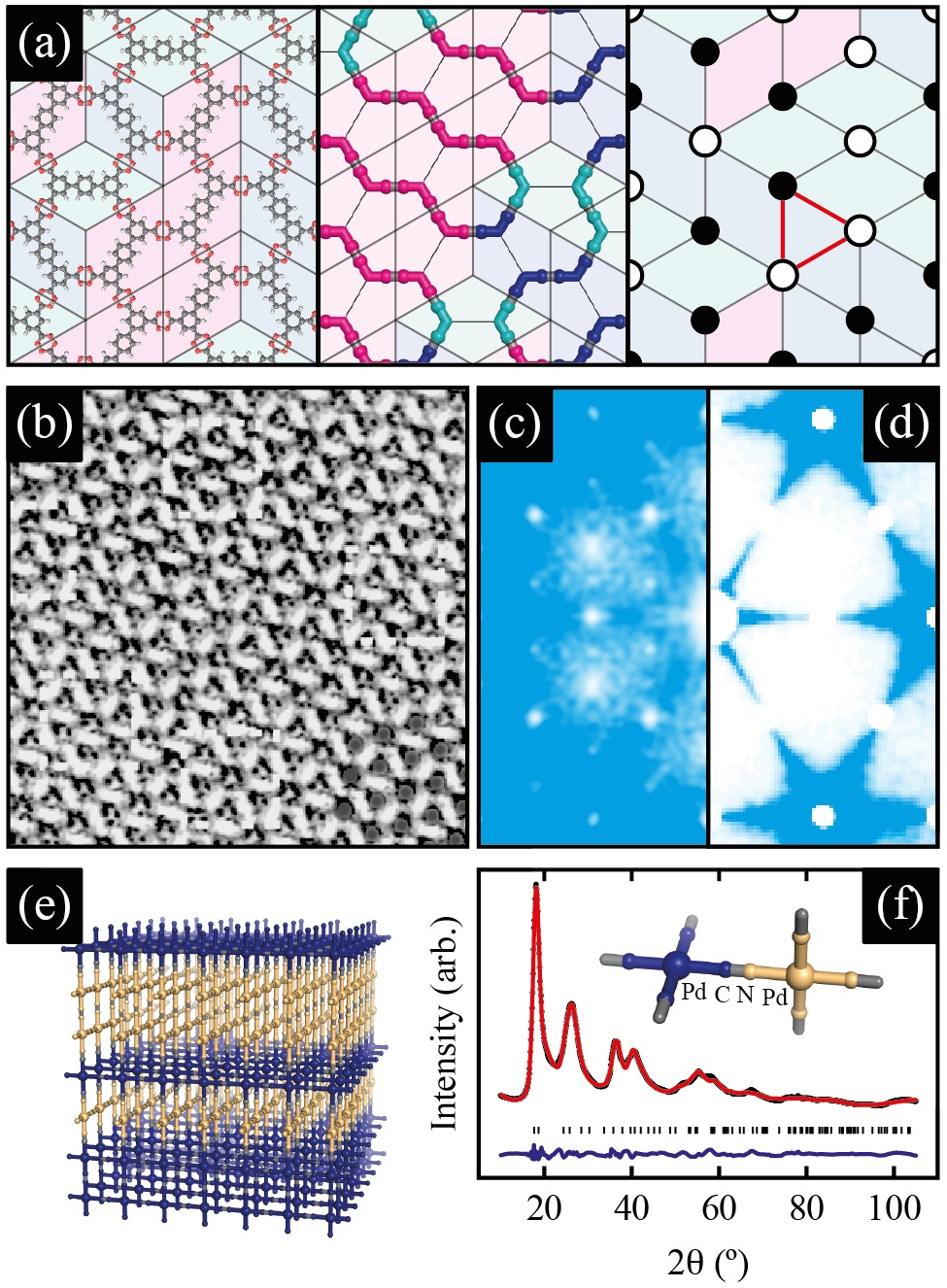}\\
FIGURE 3\\
\end{center}
\clearpage

\begin{center}
\includegraphics{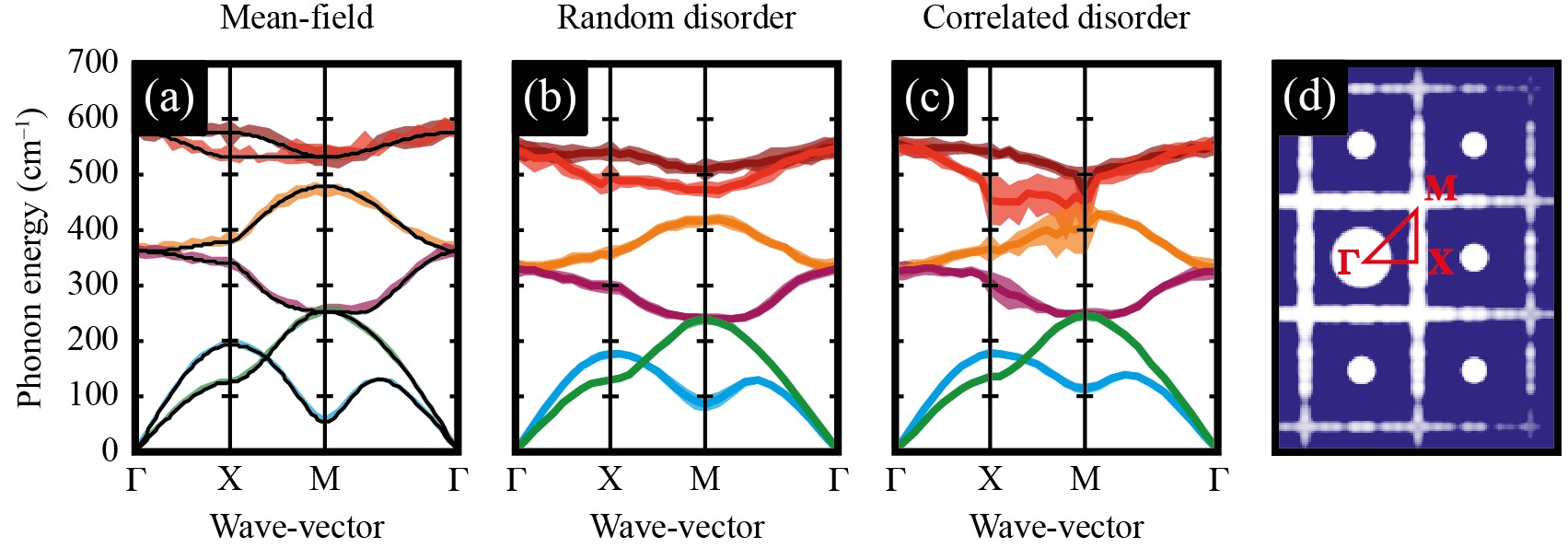}\\
FIGURE 4\\
\end{center}

\end{document}